# Magnetic Field Generation and Particle Energization at Relativistic Shear Boundaries in Collisionless Electron-Positron Plasmas

Edison Liang[1], Markus Boettcher[2] and Ian Smith[1]

[1]Rice University, MS108 6100 Main Street, Houston, TX 77005; liang@rice.edu, iansmith@rice.edu

[2]Ohio University, Physics and Astronomy Department, Athens, OH 45701; boettchm@ohio.edu

ABSTRACT

Using Particle-in-Cell simulations, we study the kinetic physics of relativistic shear flow in collisionless electron-positron (e+e-) plasmas. We find efficient magnetic field generation and particle energization at the shear boundary, driven by streaming instabilities across the shear interface and sustained by the shear flow. Nonthermal, anisotropic high-energy particles are accelerated across field lines to produce a power-law tail turning over at energies below the shear Lorentz factor. These results have important implications for the dissipation and radiation of jets in blazars, gamma-ray bursts and other relativistic outflows.

Short Title: Relativistic Shear Boundaries

Subject Keywords: Jets - Blazar; Shear Flow; Magnetic Field Generation

## 1. INTRODUCTION

An outstanding problem in modeling relativistic jets is how they can efficiently convert the outflow energy into electromagnetic turbulence, energetic particles and high-energy radiation (Mirabel & Rodreguez 2002, Boettcher 2007). While much attention has focused on shocks (Silva et al 2003, Spitkovsky 2008), the boundary layer of shear flows may constitute another important dissipation site. As the jet penetrates the ambient medium, a sharp boundary layer may be created by the large velocity difference between the jet and the ambient medium. The jet may also be accelerated to different intrinsic Lorentz factors at different distances from the axis. The resulting shear boundary is likely dissipative due to instabilities (e.g. Kelvin-Helmholtz instability (KHI), Chandrasekhar 1981). Dissipation at the shear interface of core-sheath jets offers a promising venue for relativistic particle acceleration in radio-loud AGN (Berezhko 1981, Rieger & Duffy 2006) and gamma-ray bursts (GRBs, Piran 2005). Observationally, there is also increasing evidence of a high-velocity, low-density core surrounded by a low-velocity, high-density sheath in many blazar jets. The observed limb-brightening of several VLBI radio jets is consistent with such a picture (Giroletti et al 2004). The sheath, in combination with a poloidal magnetic field, aids in stabilizing the jet propagation (Meliani & Keppens 2007, 2009, Mizuno et al 2007). Ghisellini et al. (2005) proposed a core-sheath jet as a way to overcome the Bulk Doppler Factor Crisis (BDFC) of some blazar jets (Lyutikov & Lister 2010): the rapid variability of their luminous gamma-ray emission requires large Doppler factors, in some cases exceeding 50 (Begelman et al 2008), inconsistent with the Doppler factors (10-20) inferred from VLBI radio observations. In a core-sheath jet, gamma-ray emission from the fast inner core can be more strongly beamed than the radio emission from the slower sheath, thus solving the BDFC.

In the hydrodynamic limit, shear boundary interface is unstable against the classic KHI (Chandrasekhar 1981, Drazin & Reid 1981). When ambient magnetic fields are present, a strong flow-aligned $\mathbf{B}_\parallel$ field suppresses KHI, while transverse $\mathbf{B}_T$ fields do not (Chandrasekhar 1981). Relativistic effects also reduce the KHI (Ferrari et al 1978). Gyro-kinetic simulations of space plasmas with ambient magnetic fields which give the electrons small gyro-radii, support the KHI picture, with modes unstable down to the plasma skin depth or gyroradius (Thomas & Winske 1991). However, these simulations do not address the questions of magnetic field generation (Colgate et al 2001, Medvedev & Loeb 1999) and nonthermal particle energization (Berezhko 1981, Rieger & Duffy 2006) in unmagnetized shear flows. A low-density relativistic plasma, such as those in blazar or GRB jets, is highly collisionless (i.e. Coulomb collisions are negligible, Boyd & Sanderson 2003) and needs to be modeled using Particle-in-Cell (PIC, Birdsall & Langdon 1991, BL91 hereafter) simulations. Recently, Alves et al. (2012) and Grismayer et al (2012) reported PIC simulation results of unmagnetized, low-Lorentz factor, e-ion shear flows, showing that collisionless shear boundary can create and sustain strong d.c. magnetic fields via the kinetic KHI due to fluid-like electrons with small gyroradii (Gruzinov 2008, Grismayer et al 2012). Our work differs from those of Alves et al (2012) and Grismayer et al (2012) in three major respects: (a) our shear Lorentz factors $p_o$ are much higher, (b) we focus on e+e- plasmas instead of e-ion plasmas, (c) we use 2D simulation boxes that are physically much larger than the 3D boxes used by Alves et al (2012) and Grismayer et al (2012). Using the 2.5D (2D-space, 3-momenta) LLNL code Zohar II (BL91, Langdon & Lasinski 1976), we performed simulations *separately in the shear momentum (x-y or P) plane and the transverse (y-z or T) plane*. We supplement these 2D simulations with small 3D simulations using the SNL code Quicksilver to cross check and validate the 2D results.

Our most important findings are: (a) organized quasi-stationary electromagnetic (EM) fields of alternating polarities are generated and sustained at the shear boundary by the Weibel (Weibel 1959) and 2-stream (Boyd & Sanderson 2003) instabilities, with peak magnetic fields reaching ~ equipartition values and global field energy ~ few percent of total energy; (b) nonthermal anisotropic particles are energized at the boundary layer, forming a quasi-power-law tail with low-energy turnover near the shear Lorentz factor; (c) high-energy particles are accelerated across field lines, leading to anisotropic momentum distribution and efficient synchrotron radiation; (d) the shear boundary layer exhibits a density trough due to the magnetic pressure expelling the plasma; (e) e+e- shear boundaries exhibit different properties from e-ion shear boundaries discussed in Alves et al (2012) and Grismayer et al (2012).

## 2. RESULTS

Figure 1 illustrates our 2D setup. We use 1024x2048 cells with periodic boundaries and $\sim 10^8$ particles. We also did test runs of different box sizes, cell sizes and particle numbers to ascertain that our setup gives robust and convergent answers. The initial state consists of two uniform unmagnetized electron-positron ($m_e=m_p=m$) plasmas counter-streaming with equal and opposite x-momenta $p_x/mc=+/-p_o$ in the center-of-momentum (CM) frame. We first focus on the benchmark case $p_o=15$ before comparing it to other cases. The initial temperature kT=2.5keV and particle density n=1 so that the cell size $\Delta x, \Delta y, \Delta z = c/\omega_e$=electron skin depth ($\omega_e$=electron plasma frequency). In *all figures, x, y, z are in units of $c/\omega_e$ and time is in units of $1/\omega_e$*. We use time step $\Delta t=0.1/\omega_e$ to ensure system energy conservation ($\Delta E/E_o<0.1\%$ in all runs). Due to the periodic boundaries, some particles and waves get recycled at $t\omega_e>1000$. Hence interpretation of the results at $t\omega_e>1000$ requires caution. Because of the 2D geometry, our simulation suppresses perturbation or instability in the third dimension. We refer to *2D instabilities in the x-y plane as P-modes, and those in the y-z plane as T-modes*. It turns out the two modes couple only weakly (Alves et al 2012, Karimabari et al 2012) and dominate at different times. Hence the combined 3D effects of both modes are *qualitatively* similar to the superposition of the two 2D modes (Alves et al 2012, Karimabari et al 2012).

Fig.2a shows the energy flow between particles $E_p$ and EM fields $E_{em}$ vs. time for the P-mode and T-mode. The P-mode grows rapidly and saturates early at $t\omega_e \sim 150$ with $\epsilon_B=E_{em}/(E_p+E_{em})\sim 15\%$, decaying after $t\omega_e \sim 250$ to an asymptotic value of $\sim 3\%$, while the T-mode grows slowly, saturating at $t\omega_e \sim 1000$ with $\epsilon_B \sim 13\%$ before decaying slowly to an asymptotic value of $\epsilon_B \sim 5\%$. The $\epsilon_B(t)$ value for the P-mode seems independent of the box sizes studied so far, suggesting that the boundary layer grows to a fixed fraction of the box size before field decay. However, the 13% maximum and field decay for $t\omega_e>1000$ of the T-mode are likely artifacts of the finite box size. Larger-box runs are in progress to address all scaling issues. Summing the $E_{em}$ of the two modes gives the top curve in Fig.2b, while the lower curve traces max(P,T) curves of Fig.2a. The shapes of both curves agree *qualitatively* with the result of our small 3D run (Fig.2d). The $E_{em}$ of large 3D runs likely lie between the two curves of Fig.2b with an asymptotic value of $\varepsilon_B \sim$ few % independent of box size, much higher than the saturation values of MHD results ($\varepsilon_B \sim 5 \times 10^{-3}$, Zhang et al 2009). This is expected because the MHD approximation averages out kinetic-scale fields of opposite signs. Yet kinetic-scale fields determine particle acceleration and the true emissivity of synchrotron radiation. Fig2d shows a log-linear plot of $E_{em}(t)$ to highlight its early exponential growth. Both P-mode and T-mode exhibit several "steps" due to the interactions of forward and backward propagating unstable modes (Yang et al 1994). The effective growth rates lie between $0.15\omega_e$ and $0.2\omega_e$, consistent with relativistic Weibel instability (Yoon 2007, Yang et al 1993, 1994). We check that our growth rates and fastest growing wavelengths scale as $p_o^{1/2}$ (Fig.2(c) inset), consistent with Weibel, but inconsistent with the $p_o^{3/2}$ scaling of kinetic KHI (Alves et al 2012, Grismayer et al 2012). Since $p_o \gg 1$, the e+e- pair's gyroradii are large, allowing them to freely cross the interface and interpenetrate, creating streaming instabilities at the boundary layer.

Fig.3 shows snapshots of the field profiles at sample times. Opposing particle streams crossing the shear interface generate kinetic-scale current filaments and Langmuir waves via Weibel (Weibel 1959) and 2-stream (Lapenta et al 2007, Boyd & Sanderson 2003) instabilities. Their fields grow and coalesce into larger and larger structures, eventually forming a boundary layer of several hundred skin depths, with periodic patterns of quasi-stationary magnetic fields of alternating polarities. The peak **B** fields reach equipartition values ($B^2 \sim \gamma nmc^2$), and the combined **E** fields from Weibel and 2-stream form oblique electric channels (Fig.3c). While the detailed structure and thickness of the P-mode and T-mode boundary layers appear different, qualitatively they resemble the x-y and y-z slices of small 3D simulations. Large 3D simulations will show boundary layers that combine features of the P-mode and T-mode and have thickness intermediate between the two modes. Another distinctive signature of the shear boundary layer is the density trough at the interface (Fig.4), caused by the extra magnetic pressure pushing the plasma away from the interface. The density trough created by the T-mode is deeper, wider and more persistent than the P-mode. We speculate that in large 3D runs, the density trough will be intermediate between the two modes.

We also performed parameter studies of varying $p_o$. Fig.5a compares the particle energy distributions at $t\omega_e=1000$ for different $p_o$: at low $p_o$ no power-law is formed, whereas at $p_o \geq 15$, a power law tail is evident, turning over at a $\gamma$ just below $p_o$, because field creation and accelerating the high-$\gamma$ particles drains the bulk flow energy. The power-law slope is soft due to the finite box size which recycles particles after $t\omega_e \sim 1000$. We have preliminary evidence that the power law hardens when the box size is increased, with an asymptotic slope determined by the balance between acceleration and escape from the boundary layer. This is still work in progress awaiting larger-box runs. The momentum anisotropy of the high-$\gamma$ particles also increases with $p_o$. Fig.5b compares the magnetic field evolution of the P-mode for different $p_o$: as $p_o$ increases, the field grows and saturates more slowly and are stretched more horizontally into sheet-like patterns with longer wavelengths. The boundary layer also gets thicker due to the relativistic increase of the gyroradius and skin depth.

3. SUMMARY AND DISCUSSIONS

We have demonstrated quasi-stationary field generation and particle acceleration at relativistic shear boundaries in collisionless e+e- plasmas, with local B fields reaching ~equipartition values. Particles can be accelerated across field lines to $\gamma \gg p_o$ to form power-laws. They should radiate synchrotron radiation (Rybicki & Lightman 1979, Sironi & Spitkovsky 2009ab) efficiently, turning the boundary layer into bright spots of polarized emission. Enhanced polarized radiation and density depression would be signatures of a relativistic shear boundary. Since our particle momentum distributions are anisotropic in the CM frame, additional photon beaming and Doppler boosting will result, which will not show up in imaging techniques. This may solve the BDFC of blazar jets (Lyutikov and Lister 2010). The $p_o \geq 15$ results may be relevant to GRBs: the spectra of Fig.5a for $p_o \geq 15$ resemble the generic GRB spectrum (Piran 2005). Despite local field creation, we find that the global magnetic flux is conserved to better than one part in $10^4$. Hence there is no large-scale dynamo action at the shear boundary despite the inherent vorticity, and no violation of the 2D antidynamo theorem of MHD. The e+e- shear boundary structure is fundamentally different from the e-ion shear boundary (Alves et al 2012, Grismayer et al 2013, plus our own results). In the e-ion case, the shear boundary is dominated by a monopolar slab of d.c. magnetic field supported by laminar current sheets on both sides, created and sustained by persistent ion-drift. Electrons are accelerated by charge-separation E-fields to form a narrow peak at the ion energy. But no power law tail is evident. Hence observations of shear boundary emission and structure may constrain the pair/ion ratio of relativistic jets.

This work was supported by NSF AST-0909167 and NASA Fermi Cycles 3–5. EL acknowledges the support and hospitality of LANL when part of this work was done, plus computer resource provided by LLNL and discussions with B. Langdon.

Figure Captions

Fig.1 2.5D problem set-up of the e+e- shear flow in the CM frame. The simulation boxes have $L_x=L_z=1024$ cells and $L_y=2048$ cells. The shear interfaces are located at y=512 and 1536. Circle blowup illustrates that crossing streams can lead to temperature anisotropy, instabilities and creation of EM fields. In all figures x,y,z are in units of $c/\omega_e$ and time is in unit of $1/\omega_e$.

Fig.2 (a) Time evolution of total particle energy $E_p$ and field energy $E_{em}$ for the run with $p_o=15$. Curves labeled P are the 2D results of the P-mode only and curves labeled T are the 2D results of the T-mode only. The P-mode $E_{em}$ dominates for $t\omega_e < 500$ while the T-mode $E_{em}$ dominates for $t\omega_e > 500$. The decay of T-mode $E_{em}$ at $t\omega_e>1000$ is affected by the box size. (b) The upper curve labeled P+T gives the sum of the $E_{em}$ of the P-mode and T-mode, while the lower curve labeled max(P,T) traces the maximum of the P-mode and T-mode curves of Fig.2(a). The combined contributions of both modes in 3D likely lie between these two curves. (c) Plot of $E_{em}(t)$ in a log-linear plot shows early exponential growth. The effective growth rates for B ranges ~$0.15\omega_e$ - $0.2\omega_e$, consistent with Weibel. (c) Inset compares P-mode $B_z(x)$ at y=512 at early times for different $p_o$, with dominant wavelength ~$130c/\omega_e$ for $p_o=15$ and ~ $260c/\omega_e$ for $p_o=60$. It shows that numerical Cerenkov noise (Godfrey 1974, 1975, Godfrey & Langdon 1976, Godfrey & Vay 2012, Martins et al 2009, Xu et al 2012) is kept below 10%. (d) Plot of $E_{em}(t)$ (top curve) for a small 3D run. The absolute scales cannot be directly compared to Fig.2(b), but its shape agrees qualitatively with Fig.2(b). The middle curve is magnetic energy and the bottom curve is electric energy.

Fig.3 Snapshots showing the evolution of $B_z$ (rows a & b) and $E_y$ (rows c & d) patterns for the run of Fig.2 (blue and red denote opposite signs, but color scales are different for each panel). Rows (a, c) refer to the 2D P-mode in the x-y plane. Rows (b, d) refer to the 2D T-mode in the y-z plane. The boundary layer of the T-mode is wider than that of the P-mode. Small 3D runs suggest that the 3D shear boundary structure is intermediate between the P-mode and the T-mode. y ranges from 0 to 2048. x and z range from 0 to 1024.

Fig.4 Snapshots of the density profile as functions of y, averaged over x and z, for the (a) P-mode and (b) T-mode. The T-mode density trough (row b) is deeper and wider than that of the P-mode (row a), and also persists for longer times. The counter-streaming particles diffuse in space through each other over time.

Fig.5 (a) Comparison of electron energy spectra at $t\omega_e=1000$ for different shear Lorentz factors $p_o$. From left to right: $p_o = 2, 5, 15, 30, 60$. Each spectrum peaks just below $p_o$ and gets harder for higher $p_o$. The power-law tail slope is artificially steep due to small box size. (b) Comparison of the P-mode $B_z$ pattern at two times for $p_o = 5, 30, 60$ (blue and red denote opposite polarities, but color scales are different for each panel). (x,y) ranges are the same as in Fig.3.

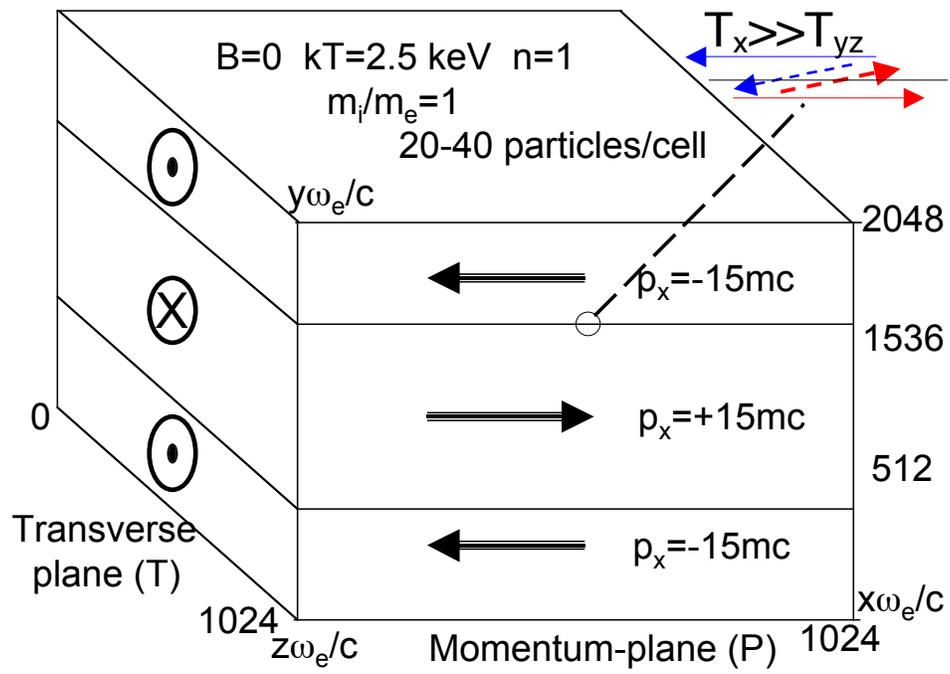

Fig.1

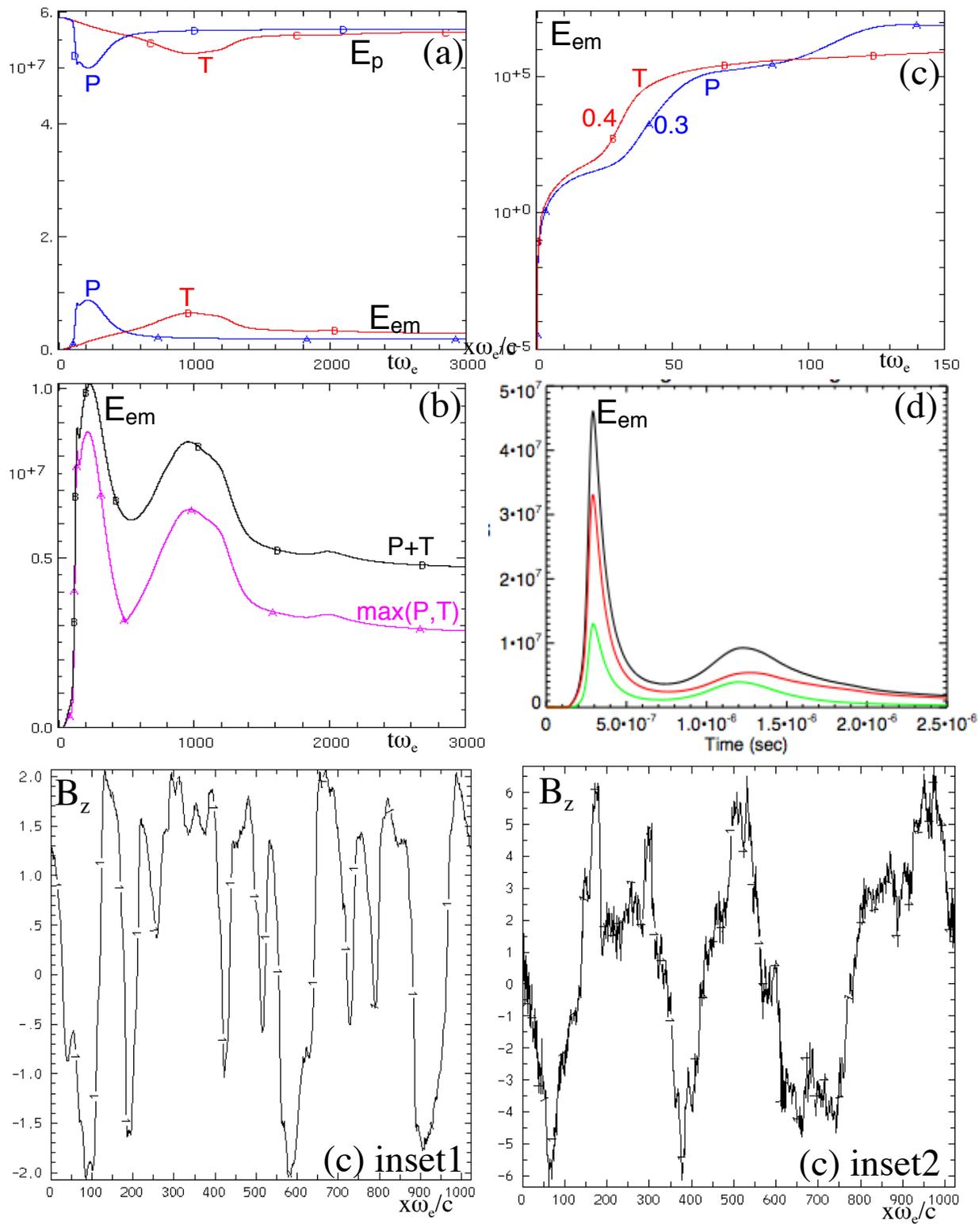

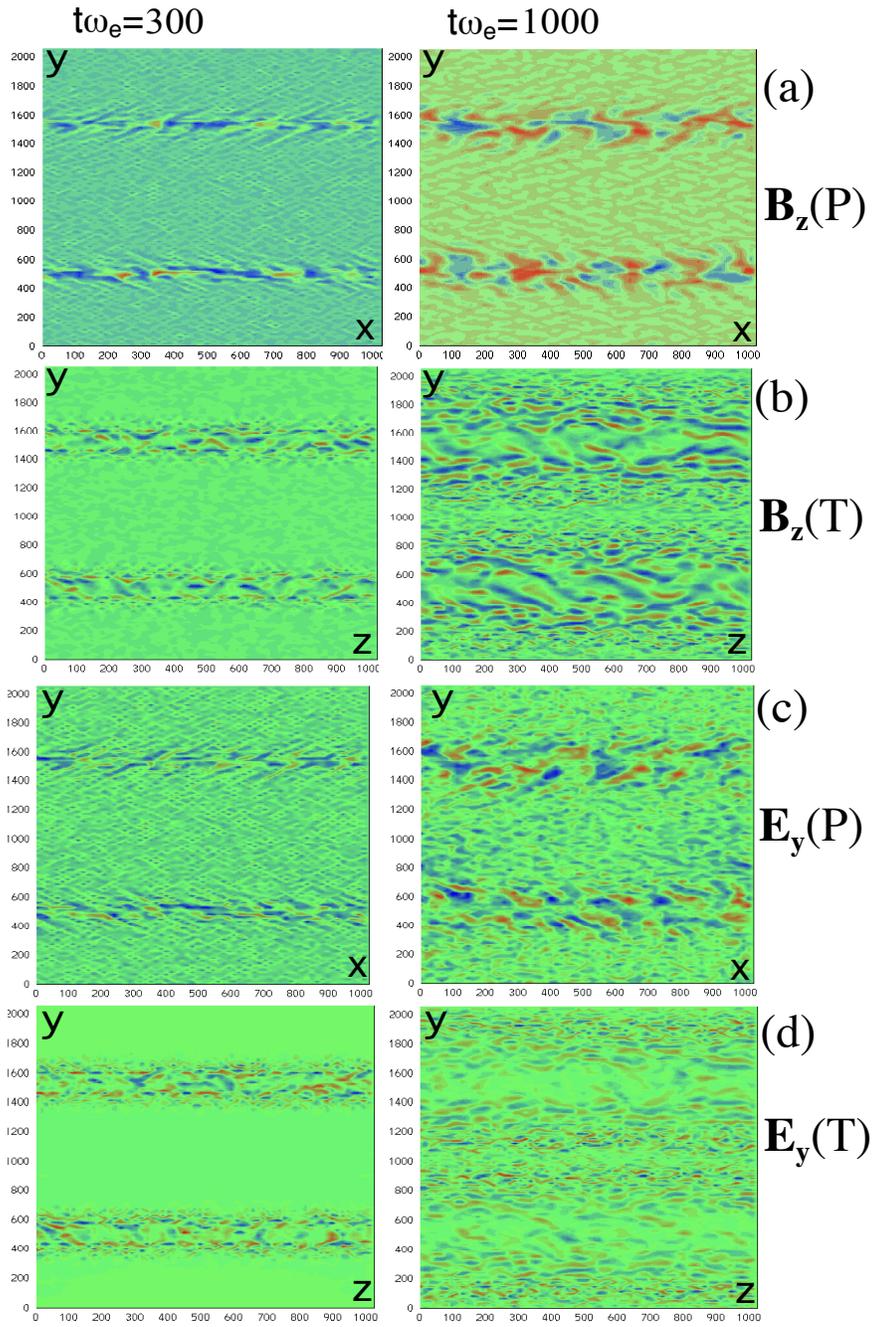

Fig.3

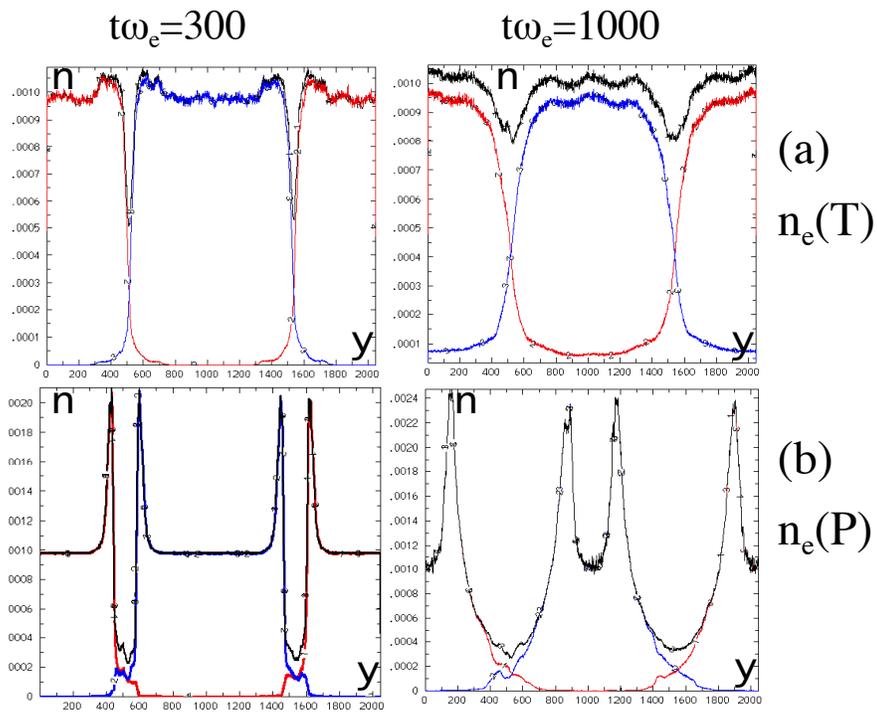

Fig.4

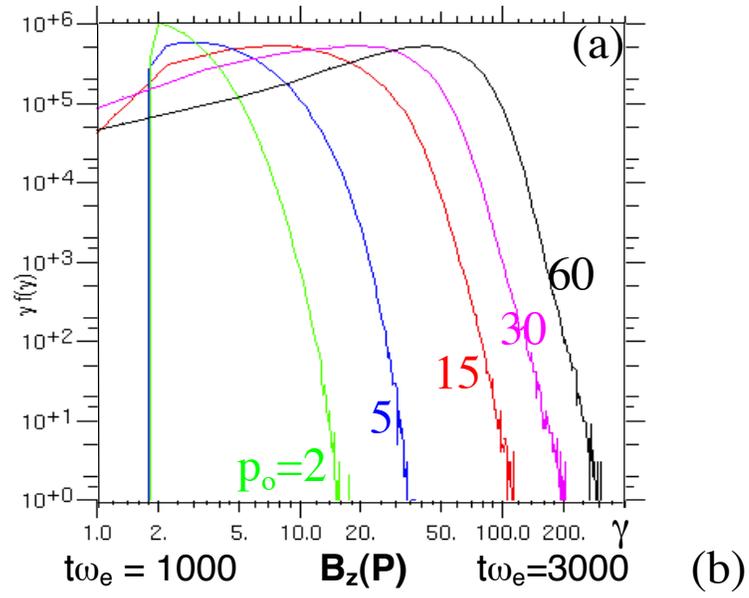

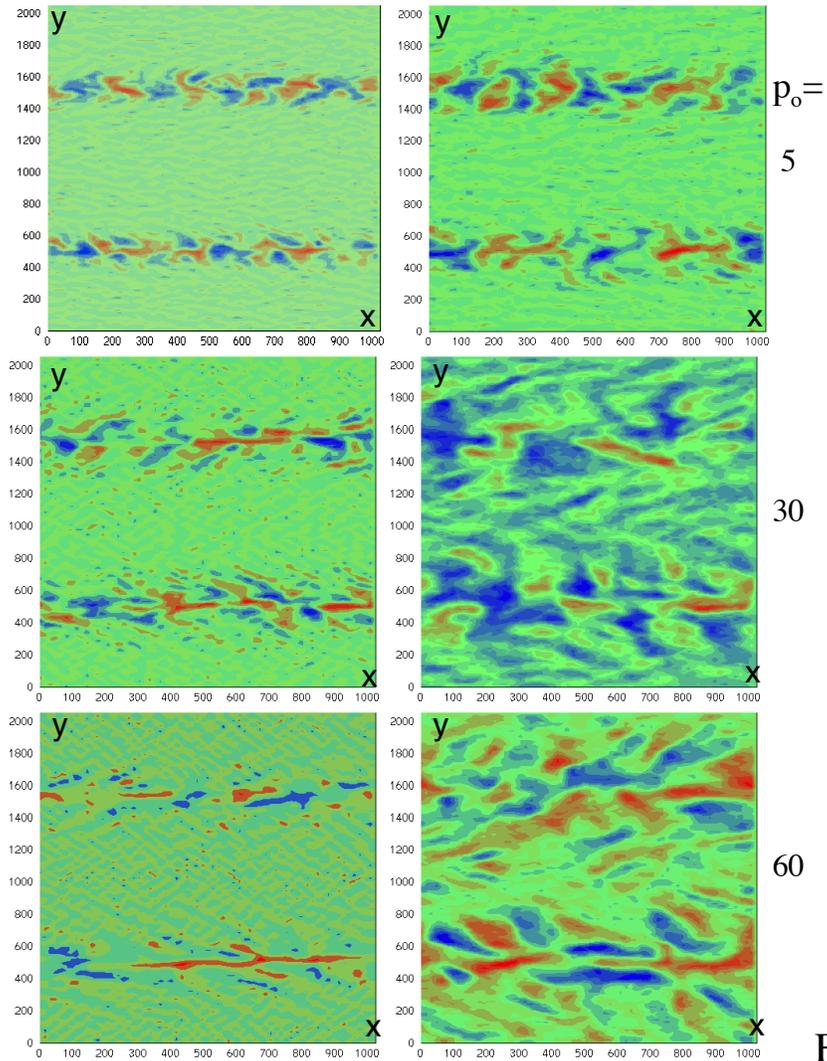

Fig.5